\documentclass[prc,twocolumn,floatfix,superscriptaddress,showpacs] {revtex4-1}
\usepackage{amsmath}
\usepackage{graphicx}
\begin{document}
\title{Time-dependent relaxation of observables in complex quantum systems}
\author{Alexander Volya}
\affiliation{Department of Physics, Florida State University, Tallahassee, FL 32306,
USA}
\author{Vladimir Zelevinsky}
\affiliation{Department of Physics and Astronomy, and National Superconducting \\
Cyclotron Laboratory/Facility for Rare Isotope Beams, \\
Michigan State University, East Lansing, MI 48824-1321, USA}
\begin{abstract}
We consider time-dependent relaxation of observables in quantum systems of chaotic and regular type. 
We show that the spread of the wave function in the Hilbert space
is determined by the survival probability which is known to have pre-exponential, exponential, and long-term power-law limiting behaviors. This result relies on complexity of the wave functions and thus is generic to many systems.   
In the chaotic limit modeled by the Gaussian Orthogonal Ensemble we show that the survival probability obtained analytically 
also fully defines the relaxation timescale of observables. This is not the case in general,  using realistic nuclear shell model 
and the quadrupole moment as an observable  we demonstrate that the relaxation time is significantly longer than defined
by  the survival probability of the initial state. An example of the non-chaotic limit of coherent and squeezed states provides an additional illustration.  
\end{abstract}
\maketitle
\section{Introduction}
The subject of thermalization in a closed quantum system of many interacting
constituents, being on a crossroad of statistical physics, quantum mechanics,
condensed matter and nuclear physics, attracts currently a great interest,
both theoretical and experimental, see recent review papers \cite{polkovnikov,BISZ,mori18}.
There are several deep reasons for this interest. First of all, such systems are currently
on the frontiers of experimental research. Cold atoms and molecules in traps, nanodevices,
interacting spin systems etc. are studied worldwide and being applied to diverse purposes,
including broad technical applications. Future quantum computers by necessity will belong
to this class as their elements should interact on a microscopic scale in order to be able
to produce practical results. Complex atoms, molecules (including biological), and atomic nuclei
by their nature are systems of interacting quantum constituents.
In many cases, such as an isolated atom, nucleus, or molecule, the system is self-bound
and lives in its intrinsic stationary state without any external heat bath prior to its use in
an experiment. An important theoretical and practical question is if such a system, in spite
of a relatively small number of degrees of freedom, can be described by the standard application
of statistical ideas not referring to the {\sl thermodynamic limit} of macroscopic volume and
large number of particles. If the answer could be positive, the treatment of the system and
of many applications would be significantly simplified.
This question is closely tied to the fundamental issues of quantum mechanics such as linearity,
decoherence, and hidden degrees of freedom \cite{weinberg:1989}. Formally, it is known that 
the dynamics of wave function components in a  closed system with a finite Hilbert space is subject 
to classical equations of motion and is quasiperiodic.  It has been shown that ergodicity 
and canonical ensemble resembling thermodynamics can be introduced with non-linear terms 
\cite{kusnezov:1995,kusnezov:1993,bulgac:1990} making Hilbert space dynamics chaotic.
Effectively, a system with high level density and sufficiently complicated Hamiltonian develops 
intrinsic chaoticity that eliminates the need in an external heat bath.
A serious argument in favor of a statistical approach is related to the understanding of
the phenomena of quantum chaos \cite{porter,brody}. There is no need to directly require that
the behavior of a realistic quantum system coincide with that of random matrix ensembles.
At a sufficiently high density of stationary states, their wave functions should be universally
mixed in each class with the fixed values of exact constants of motion. It was qualitatively
understood long ago \cite{parceval} and formulated as a qualitative statement that all typical
wave functions in the same energy region of  such a complex system ``{\sl look the same}".
Moreover, as it was indicated even much earlier in {\sl Statistical Physics} by Landau and
Lifshitz \cite{LLSF}, the observables found for such quantum states are essentially the same as
for the equilibrium statistical ensemble, just expressed in terms of energy rather than of
temperature. Nowadays, such statements are widely accepted being named the {\sl eigenfunction
thermalization hypothesis} (ETH \cite{ETH}).
Reliable illustrations to those statements come from realistic atomic \cite{grib} and nuclear
\cite{big} calculations. In both cases we deal with a self-bound system of interacting particles,
while the interaction matrix elements can be conveniently expressed in the mean field basis
that supposedly incorporates the regular (non-chaotic) features of the dynamics. The exact
solution of the many-body quantum problem in the finite orbital space reveals the internally
developed chaotic behavior of stationary states and observables starting already at a moderately
high excitation energy. This behavior, with a smooth energy dependence, can be naturally
translated into thermodynamic language with effective temperature and entropy. It is important
that the Hamiltonian of a system does not contain any random elements $-$  the system is
effectively thermalized by intrinsic interactions.
A special interest leads us to the non-stationary development in isolated quantum systems.
We will not review here an enormous literature devoted to the comparison of non-equilibrium
classical and quantum dynamics. We appraise the statement from \cite{mori18} that
``{\sl In physical terms, one may say that quantum thermalization occurs in the Hilbert space
rather than phase space}." There are recent conjectures \cite{maldacena15} of a universal
bound for the speed of chaotization of the system generalizing the classical Lyapunov exponent.
Our instruments below will be the limiting case of mathematical quantum chaos and the nuclear
 shell model as a realistic description of the set of stationary states in an isolated interacting 
many-body quantum system. Similarly to the approach used in classical Hamiltonian dynamics 
we will directly study the time evolution of the quantum wave packets in the ideal situation
and in a realistic case.
This work pursues  a broad goal of understanding how non-stationary states and perturbations in
a complex many-body systems evolve in time, what are the mechanisms for thermalization and decoherence,
what role complex interactions and mean field play in this dynamics. Finally we seek a connection
to the physics of quantum decay, both exponential and non-exponential \cite{decay}.
To explore these questions we consider a series of models. We include a summary of well known results
describing the behavior of coherent and squeezed states of a quantum harmonic oscillator which we put
forward as an example of the non-chaotic (regular) limit. Then we formulate questions and present a detailed
analytical and numerical study of time evolution of states and perturbations in the
Gaussian Orthogonal Ensemble (GOE) considered as an opposite extreme, a completely random quantum system.
As intermediate examples we consider a bosonic system with two-body random interactions and the realistic
description of the $^{24}$Mg nucleus with the abundant history of the shell-model studies in the limited but
still sufficiently rich orbital space of twelve valence particles (six protons and six neutrons above the inert
core of the doubly-magic $^{16}$O). The specific choice of the target and
nuclear quadrupole moment $Q$ as an observable is related to its clear classical analog with well
known empirical manifestations of trends to deformation in this nucleus \cite{prit}. The eigenstates
of such an operator serving as an analog of a coordinate-localized classical state are not stationary
and evolve as quantum superpositions. In some respect, this evolution is the road from a simple to
compound state mixing many degrees of freedom in a quantum system.
\section{Quantum evolution}
\subsection{Starting point}
We consider a closed  quantum system with complete (in our case finite but practically quite large)
set of stationary states $|n\rangle$ normalized according to $\langle n|n'\rangle=\delta_{nn'}$
and their energies $E_{n},$ eigenvalues of the time-independent Hermitian Hamiltonian $\hat{H}$
\begin{equation}
\hat{H}|n\rangle = E_{n} |n\rangle.
\label{eq:1}
\end{equation}
At the initial time $t=0$ we form a normalized wave packet,
\begin{equation}
|\Psi_{a}\rangle=\sum_{n}a_{n}|n\rangle,
\label{eq:2}
\end{equation}
In particular, the initial state might be an eigenstate of some operator $\hat{Q}$ describing a generalized
coordinate of the system after this variable has been measured by some device. Another choice could be
an eigenstate of a non-interacting many-body system with the Hamiltonian $\hat{H}_{0}$ so that the
evolution would describe the behavior after the  interaction is suddenly turned on,
 $\hat{H}=\hat{H_{0}}+\lambda \hat{V}.$
The time evolution of the initial
state, expressed in terms of eigenstates, is just a phase dynamics given by ($\hbar=1$)
\begin{equation}
a_{n}(t)=a_{n}e^{-iE_{n}t}.
\label{eq:3}
\end{equation}
Obviously, the overlaps of two initial packets $\langle \Psi_{a}(t)|\Psi_{b}(t)\rangle=\langle \Psi_{a}(0)|\Psi_{b}(0)\rangle$
do not depend on time although both packets evolve and spread. In order to see the physical evolution,
there are many choices (one particularly interesting case is the Loschmidt echo \cite{gorin06}). The expectation value
of a simple physical variable $\hat{Q}$ changes according to
\begin{equation}
Q_{a}(t)=\langle \Psi_{a}(t)|\hat{Q}|\Psi_{a}(t)\rangle.
\label{eq:4}
\end{equation}
It can be effective to discuss the same physics using the Heisenberg picture of time-dependent operators where
$Q_{a}(t)=\langle \Psi_{a}(0)|\hat{Q}(t)|\Psi_{a}(0)\rangle$ and
\begin{equation}
\hat{Q}(t)=e^{i\hat{H}t}\hat{Q} e^{-i\hat{H}t}=\sum_{n=0}^{\infty} \frac{(it)^n}{n!} [\hat{H},[\hat{H},\dots,[\hat{H},\hat{Q}\underbrace{]]\dots]}_n.
\label{eq:5}
\end{equation}
To associate the width of quantum packets with classical phase space dynamics the uncertainty
\begin{equation}
\langle\Delta \hat{Q}\rangle^2_a \equiv {\langle \Psi_{a}|\hat{Q}^2|\Psi_{a}\rangle}- ({\langle \Psi_{a}|\hat{Q}|\Psi_{a}\rangle})^2
\label{eq:6}
\end{equation}
is introduced.
In order to see the dynamics of conjugate variables, an analog of the classical phase space and symplectic structure of
Hamiltonian dynamics, we consider the time derivative of the ``coordinate" $Q$, an operator $\hat{P}(t)=\hat{\dot{Q}}(t)$,
\begin{equation}
\hat{P}=-i[\hat{H},\hat{Q}].                                           \label{eq:7}
\end{equation}
\subsection{Coherent state}
To introduce a point of reference, we briefly show the case opposite to chaos, namely an example of
special coherence where the dynamics is obviously absolutely regular. A similar case of squeezed states
was discussed in \cite{moeckel09}. For a simple harmonic oscillator of mass $m=1$ and frequency $\omega$,
we take the variable as a coordinate expressed in terms of creation, $\hat{c}^{\dagger}$, and annihilation,
$\hat{c}$, operators,
\begin{equation}
\hat{Q}=\,\frac{1}{\sqrt{2\omega}}\,(\hat{c}+\hat{c}^{\dagger}).                \label{eq:8}
\end{equation}
The Hamiltonian here is
\begin{equation}
\hat{H}=\omega\left (\hat{c}^\dagger \hat{c}^{\dagger}+\frac{1}{2}\right ).                \label{eq:9}
\end{equation}
The eigenstates are those with a certain number of quanta,
\begin{equation}
|n\rangle=\,\frac{(\hat{c}^{\dagger})^{n}}{\sqrt{n!}}|0\rangle,                        \label{eq:10}
\end{equation}
Considering the evolution of a coherent state $|\Psi_{\alpha}\rangle$ whose expansion coefficients in the eigenbasis are
\begin{equation}
a_{n}=e^{-|\alpha|^{2}/2}\,\frac{\alpha^{n}}{\sqrt{n!}},                            \label{eq:11}
\end{equation}
we note that time dependence of $a_n(t)$ amounts to a phase dependence of $\alpha(t)=\alpha e^{-i\omega t}.$
As coherent states are eigenstates of the annihilation operator, $\hat{c}|\Psi_{\alpha}\rangle =\alpha |\Psi_{\alpha}\rangle$,
we come to
\begin{equation}
Q(t)=\sqrt{\frac{2}{\omega}}\,{\rm Re}\left (\alpha e^{-i\omega t} \right )\,, \quad
P(t)= \sqrt{{2\omega}}\,{\rm Im}\left (\alpha e^{-i\omega t} \right ).
 \label{eq:12}
\end{equation}
By the original design, being eigenstates of the annihilation operator,
the coherent states minimize the uncertainty relation and give equal uncertainty to both momentum and coordinate
\begin{equation}
\langle\Delta \hat{Q}\rangle^2_{\alpha}=\langle\Delta \hat{P}\rangle^2_{\alpha}=\frac{1}{2}
\label{eq:13}.
\end{equation}
Eq. \eqref{eq:13} remains valid in the course of time evolution while the state remains a coherent state.
The dynamics of $Q$ and $P$ in coherent states representing Gaussian wave packets is fully equivalent to classical
oscillatory motion.
While for different initial states the uncertainties can be different the result in eq.  \eqref{eq:12} is general
as the commutation relations allow one to solve the Heisenberg equations of motion in a closed form; the same result
follows also from \eqref{eq:5}. The standard normalization and orthogonality relations for coherent states determine
the oscillating survival probability
\begin{equation}
\langle \Psi_{\alpha}(t)| \Psi_{\alpha}(0)\rangle =\exp\left (-4\alpha^2 \sin^2(\omega t/2)\right).
\label{eq:14}
\end{equation}
Another specific example worth mentioning is the evolution of the state $|a\rangle$ prepared as an eigenstate of
the coordinate operator, $\hat{Q} |a\rangle = a|a\rangle$. This state belongs to the
group of squeezed states with the infinitely large squeezing parameter that gives the infinite uncertainty to momentum and
zero to coordinate,
\begin{equation}
|a\rangle = \left(\frac{\omega}{\pi}\right )^{1/4}\, \exp\left (-\frac{\omega}{2}\, a^2\right )
\exp \left (\sqrt{2\omega} a \hat{c}^\dagger-\frac{1}{2}\hat{c}^\dagger \hat{c}^\dagger
\right ) |0\rangle.
\label{eq:15}
\end{equation}
The expansion of these states in the stationary basis
of the harmonic oscillator is obvious in terms of Hermite polynomials.
The time evolution of coordinate and momentum in the squeezed state still follow the same classical equations. The uncertainty
in coordinate and momentum can be
represented by the phase-space ellipse (here infinitely squeezed in coordinate)
that rotates as a function of time along with the oscillation of the center.
The initial wave function sharply localized in coordinate has infinitely small width and in the course of time evolution its
survival probability instantaneously goes to zero.
\section{Chaotic dynamics}
\subsection{Universal limit}
Here we assume that the amplitudes $a_n$ in eq. (2) are uncorrelated Gaussian variables for the majority of possible choices
of the states $|\Psi_{a}\rangle$ (an arbitrary orientation of the state vector in the multidimensional Hilbert
space).
This rather weak condition is almost always satisfied in realistic many-body systems of interacting particles, at least
starting from some excitation energy.
This results in the Porter-Thomas distribution for spectroscopic factors (squares of amplitudes).
Without any preference for a particular orientation in the Hilbert space it is appropriate to view the set of
stationary states $|n\rangle$ as just one of many choices of basis states providing a coordinate system in the multidimensional space.
Apart from being eigenstates of the Hamiltonian with the trivial time dependence, the set of stationary states is not any different by its statistics
from a system of states $|a\rangle$ built as eigenstates of some other Hermitian operator $\hat{Q}$ that is not a constant of motion.
Then it makes sense to perform averaging over the ensemble of components which are uncorrelated and have normal distribution so that
\begin{equation}
\overline{\langle a|n \rangle}=0\,
\label{eq:16}
\end{equation}
and
\begin{equation}
\overline{\langle a|n \rangle \langle n'|a' \rangle}=\frac{1}{{\cal N}}\delta_{a a'} \delta_{n n'}.
\label{eq:17}
\end{equation}
Under the assumption of orthogonal invariance, the characteristic number of components, ${\cal N}$,  should be approximately the
same for the majority of choices of $\hat{Q}$ so we can view  ${\cal N}$ as being a constant and representing the effective
dimension of the Hilbert space. In this consideration the operator $\hat{H}$ is special only by its role as a generator of time
development.
The matrix elements of the propagator,
\begin{equation}
f_{a'a}(t)=\langle a'| e^{-i\hat{H}t}|a\rangle\,,
\label{eq:18}
\end{equation}
subject to the self-evident conditions
\begin{equation}
f_{aa'}(0)=\delta_{aa'}\,,\quad f_{aa'}(t)=f_{a'a}(-t),                        \label{eq:19}
\end{equation}
play a special role in the following discussion.
Similarly to the studies of fidelity \cite{gorin06,KTS98}, the squares of these amplitudes represent
the density matrix.
From eq. \eqref{eq:17} it immediately follows that, in average, only diagonal matrix elements of the propagator are not vanishing,
\begin{equation}
\overline{f_{aa'}(t)}=\sum_{n} e^{-i E_n t} \overline{\langle a|n \rangle \langle n|a' \rangle} =  \delta_{aa'} f(t),
\label{eq:20}
\end{equation}
where we define $f(t)\equiv \overline{f_{aa}(t)}.$
The additional averaging
over all states $a$ leads to the trace of the Hamiltonian matrix,
\begin{equation}
F(t)=\frac{1}{{\cal N}} \sum_n  e^{-iE_n t}=\frac{1}{\cal N} \, {\rm Tr} (e^{-i\hat {H}t}).
\label{eq:21}
\end{equation}
Under the above assumptions of full randomness, the
function $f(t)$ is universal; it is the same for any complex state $|a\rangle$ and
$f(t)=F(t).$ The diagonal survival amplitude $f_{aa}(t)$ is a macroscopic function so its variance is of the order 
of $1/{\cal N}$ allowing us to assume that  $\overline{|f_{aa}(t)|^2}=|f(t)|^2,$ and similarly replace averages 
of products with products of averages elsewhere.
The set of ${\cal N}-1$ complex numbers $f_{a}(t)\equiv f_{a a_0}(t)$, where
$a\ne a_0$, for any initial state $a_0$ is statistically equivalent
which allows us to suppress the subscript.
The amplitudes $f_a(t)$ are random, normally distributed components of an evolved vector $e^{-i\hat{H}t}|a_0\rangle$ projected onto
the space orthogonal to the original vector $|a_0\rangle.$
From the normalization
\begin{equation}
|f(t)|^2+\sum \overline{|f_a(t)|^2}=1
\label{eq:22}
\end{equation}
we obtain the variance of the distribution
\begin{equation}
\overline{|f_a(t)|^2} =  \frac{1-|f(t)|^2}{{\cal N}-1}.
\label{eq:23}
\end{equation}
The matrix elements $f_a(t)$ are complex and both real and imaginary parts are subject to normal distribution, although the widths of real and imaginary parts are not generally equal.
Suppose that an initial state is prepared by a measurement of a Hermitian operator $\hat{Q}$ that sets its value
to $Q_a$.
We assume that the states
$|a\rangle$ form a complete orthonormal set of the eigenstates of $\hat{Q}.$
Separating the diagonal and off-diagonal components, the introduced statistics lead to the time evolution
\begin{equation}
Q_a(t)=\sum_{a'} Q_{a'} |f_{a'a}(t)|^2= Q_a |f(t)|^2  + \frac{1-|f(t)|^2}{{\cal N}-1} \sum_{a'\ne a} Q_{a'}.
\label{eq:24}
\end{equation}
Without loss of generality, we can
define
${\rm Tr}\, \hat{Q}=\sum_{a'} Q_{a'}=0$ 
and, in the same way, ${\rm Tr} \hat{H}=\sum_n E_n=0.$ With ${\cal N}\gg1$.
the contribution of a single term $a'=a$ is not significant and we can view the sum as representing the
average over
all possible values of $Q.$ The following discussion of this section is carried out under these assumptions along with the condition ${\cal N}\gg1.$ Further in this work we address limitations of this approach.
Thus, for any initial eigenstate of the coordinate $\hat{Q}$, the time evolution is universal
\begin{equation}
Q(t)= Q(0) |f(t)|^2.
\label{eq:26}
\end{equation}
Following similar steps, the mean square fluctuation \eqref{eq:13}
can be evaluated as
\begin{equation}
\langle\Delta \hat{Q}\rangle^2_a(t)=Q_a^{2}(|f|^{2}-|f|^{4})+(1-|f|^2)\frac{1}{{\cal N}} \sum_{a'} Q_{a'}^{2}.                            \label{eq:27}
\end{equation}
As $|a\rangle$ is an eigenstate of $\hat{Q}$, at the initial moment $\langle\Delta \hat{Q}\rangle^2_a=0.$
If the energy spectrum is symmetric,
 the function $f(t)$ is real as $f^*(t)=f(-t).$
For the time derivative in eq. \eqref{eq:7} it is convenient to introduce the energy scale $\lambda$
and consider the dimensional time $\tau=\lambda t$ thus having the same units for $Q$ and $P.$
Then the average diagonal component of the time derivative of the propagator is
\begin{equation}
\frac{d}{d\tau} \overline{ \langle a|e^{- i \hat{H}t}|a\rangle }=-i\overline{\langle a|{(\hat{H}}/\lambda) e^{- i (\hat{H}/\lambda) \tau}|a\rangle } =  f'(\tau).
\label{eq:28}
\end{equation}
The off-diagonal components have a random distribution with the mean value
\begin{equation}
\frac{1}{\lambda^2}\overline{|\langle a' |\hat{H} e^{- i \hat{H}\tau/\lambda}|a\rangle|^2 } = \frac{1-|f'(\tau)|^2}{\cal N}.
\label{eq:29}
\end{equation}
We select the normalization of the scaling variable  $\lambda$ so that it represents the spectral width
\begin{equation}
\lambda^2=\frac{1}{\cal N} {\rm Tr} \hat{H}^2-\left (\frac{1}{\cal N}{\rm Tr} \hat{H} \right )^2.
\label{eq:30}
\end{equation}
For the momentum operator, eq. \eqref{eq:7}, only the diagonal part of the evolution operator contributes,
\begin{equation}
P(\tau)=\frac{d}{d\tau}  Q(\tau)=  Q(0)\, (f^* f'+{f'}^* f).
\label{eq:31}
\end{equation}
Similarly to the coordinate, the momentum time-dependence for individual eigenstates $a$ of the operator $\hat{Q}$ is also universal. Initially, $P(0)=0$ because the initial decay rate of unstable states is always zero \cite{decay}.
The average square of the momentum operator involves off-diagonal terms,
\begin{equation}
\langle a(\tau) |P^2| a(\tau) \rangle  =Q^2_a \left (|f|^2+{|f'|}^2 \right ) + (2-|f|^2-|f'|^2) \frac{1}{{\cal N}}
{\rm Tr}{\hat{Q}}^2.
\label{eq:32}
\end{equation}
\subsection{GOE case}
\begin{figure}[h]
\begin{center}
\includegraphics[width=0.99\linewidth]{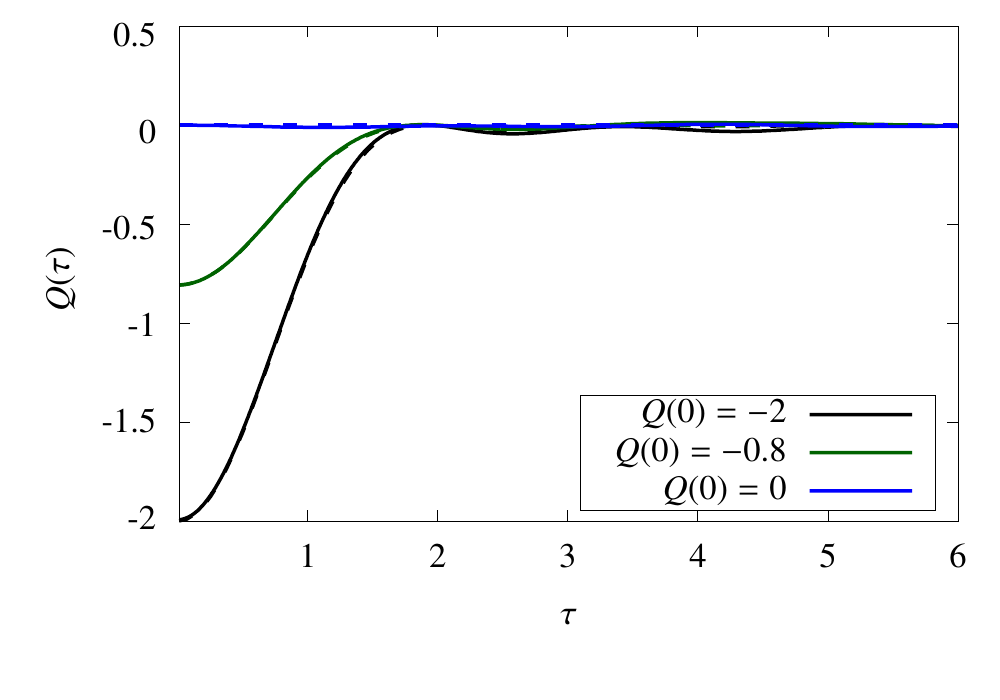}
\end{center}
\caption{\label{fig:goe_evo_q} Time evolution of
of the variable $Q(\tau).$ Numerical (solid lines) and nearly indistinguishable analytical results (dashed lines) are shown.
}
\end{figure}

\begin{figure}[h]
\begin{center}
\includegraphics[width=0.95\linewidth]{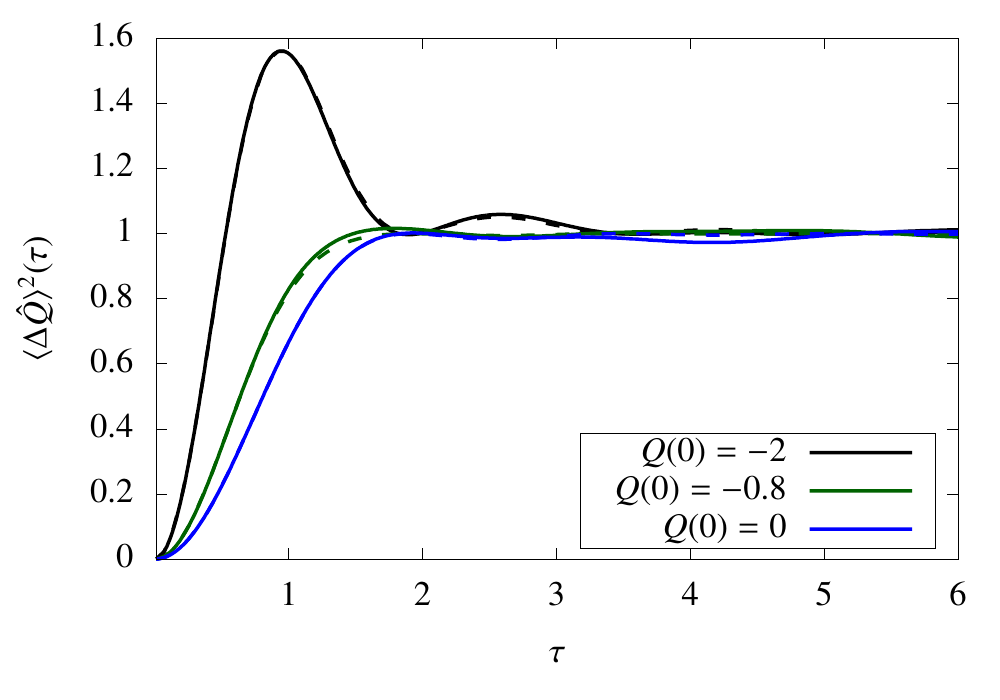}
\end{center}
\caption{\label{fig:goe_evo_Dq} Time evolution of  $\langle\Delta \hat{Q}\rangle^2(\tau).$ Numerical  and analytical results are shown.
}
\end{figure}

\begin{figure}[h]
\begin{center}
\includegraphics[width=0.95\linewidth]{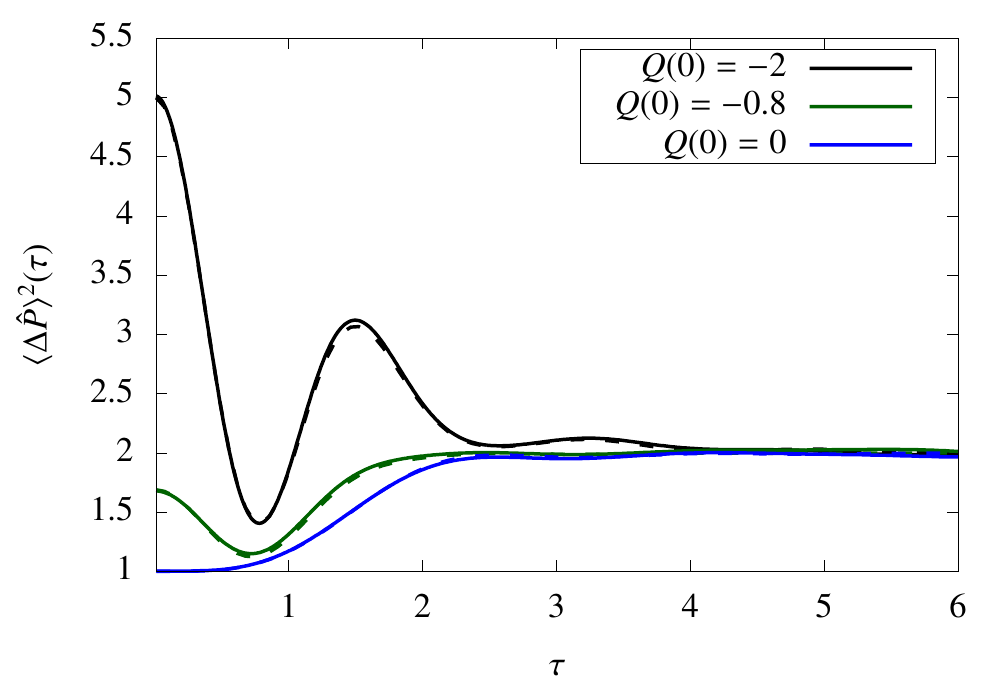}
\end{center}
\caption{\label{fig:goe_evo_Dp} Time evolution of $\langle\Delta \hat{P}\rangle^2(\tau).$ Numerical and analytical results are shown.
}
\end{figure}

\begin{figure}[h]
\begin{center}
\includegraphics[width=0.95\linewidth]{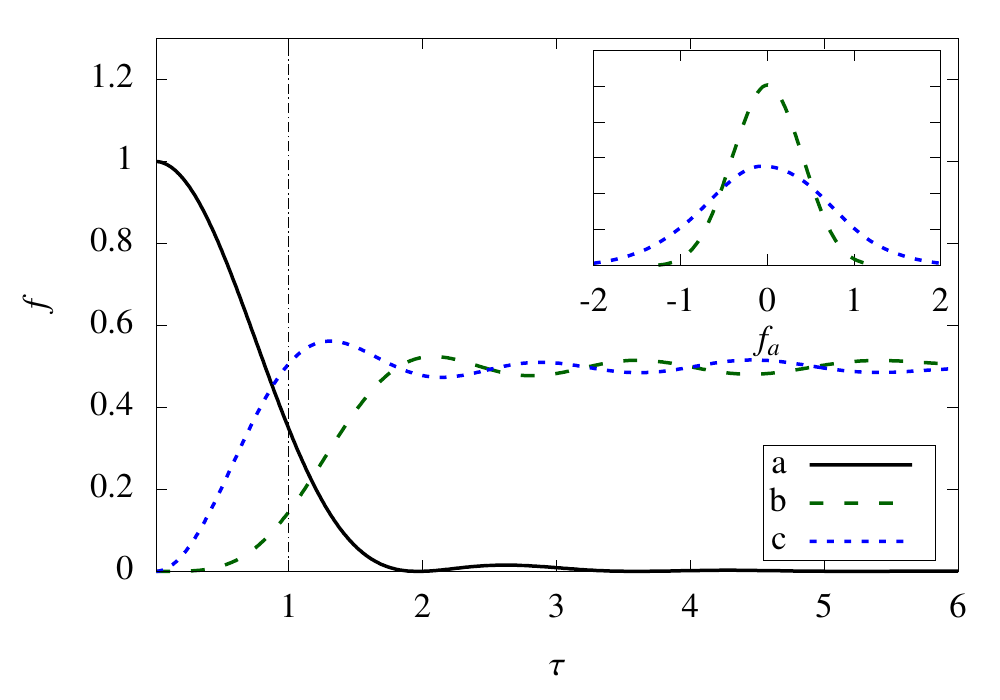}
\end{center}
\caption{\label{goe_evo_gab_w} (a) Components of the sum (\ref{eq:38}),  $|f(t)|^2;$  (b) ${\cal N} \overline 
{{f^{(r)2}_a}}$;  (c) ${\cal N} \overline {{f^{(i)2}_a}}$.
}
\end{figure}

\begin{figure}[h]
\begin{center}
\includegraphics[width=0.95\linewidth]{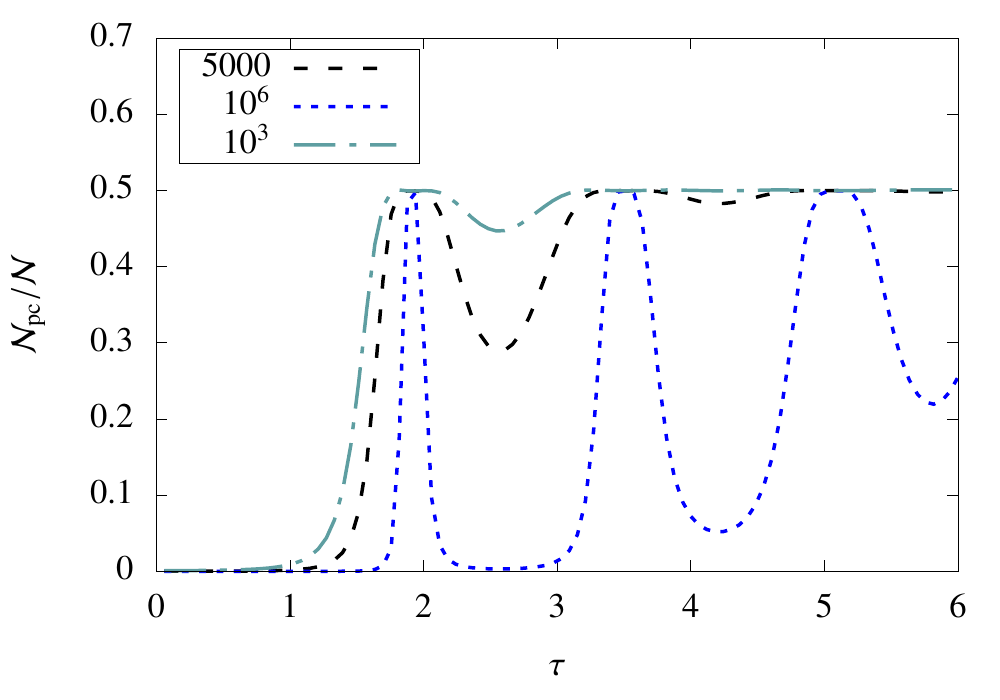}
\end{center}
\caption{\label{goe_evo_NPC_M} The ${\cal N}_{\rm pc}/{\cal N}$ as a function of time, eq. \eqref{eq:42}.}
\end{figure}

\begin{figure}[h]
\begin{center}
\includegraphics[width=0.95\linewidth]{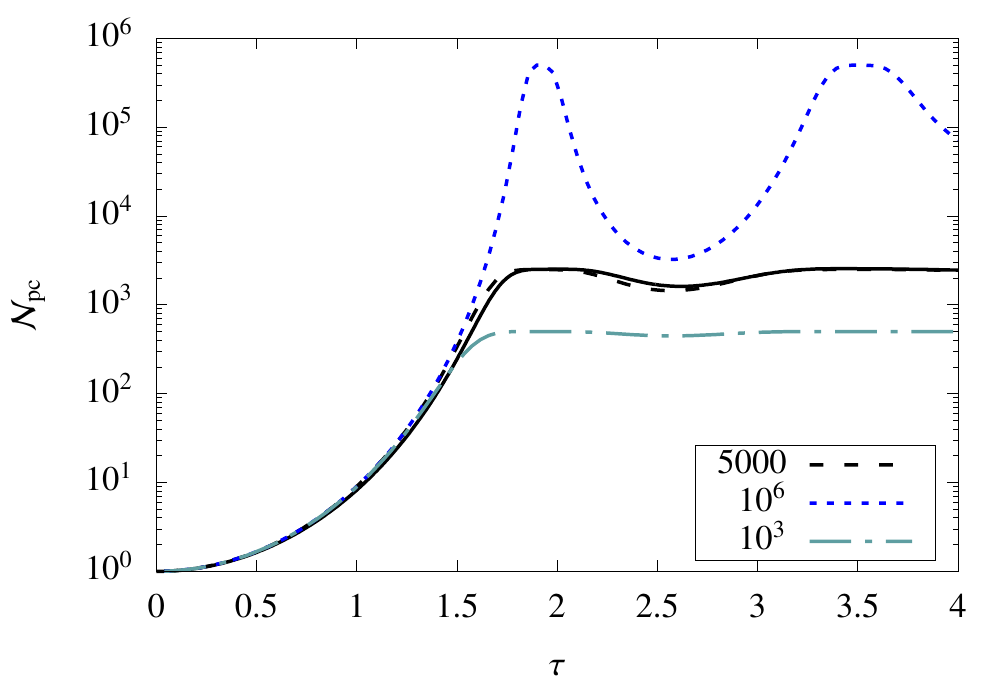}
\end{center}
\caption{\label{goe_evo_NPC} The ${\cal N}_{\rm pc}$ as a function of time. eq. \eqref{eq:42}  is used for all dashed lines, 
the numerical result for ${\cal N}=5000$ is shown with a solid line to highlight agreement.
}
\end{figure}

As a limiting case of chaotic dynamics we
take the Hamiltonian modeled by the GOE; in numerical examples the matrix dimension is $10^{4}$.
The ensemble  level density is given by a semicircle of radius $R=2\lambda$, while the parameter $\lambda=1$.
The energy scale is determined by the variance $\lambda=R/2$.
Due to the orthogonal invariance, we use an eigenbasis of
the generalized coordinate $Q$.
 The results are expressed in terms of the Bessel functions,
\begin{equation}
f(\tau)= \frac{1}{2\pi }\int_{-2}^{2}d\epsilon\,\sqrt{4-\epsilon^{2}}\,e^{-i \epsilon\tau}=
\frac{1}{\tau} J_1 (2\tau),
\label{eq:33}
\end{equation}
\begin{equation}
f'(\tau)=-\frac{2}{\tau}\, J_2 (2\tau).
\label{eq:34}
\end{equation}
Figures \ref{fig:goe_evo_q}, \ref{fig:goe_evo_Dq} and \ref{fig:goe_evo_Dp} show the evolution
$Q(\tau),$ $\langle\Delta \hat{Q}\rangle^2(\tau),$ and $\langle\Delta \hat{P}\rangle^2(\tau),$ respectively, for several initial 
eigenstates of the operator
$\hat{Q}$ with different values $Q(0).$ The dashed lines in figures show analytical results from the previous section, which, being nearly indistinguishable from the numerics, confirm those arguments. The evolution consists of the prethermalization
dependent on the choice of the initial state and universal thermaized stage with almost identical small fluctuations.
In all numerical studies we use a single realization of the GOE without any averaging, and the quality of agreement shows that the statistical assumptions are valid for each individual realization.
The agreement improves for higher space dimensions.
The {\sl number of principal components}, ${\cal N}_{{\rm pc}}$, is a commonly discussed quantity characterizing the complexity of the wave function and its spreading degree with respect to a certain basis. In realistic many-body systems, this
can be a mean-field basis (for example, a Slater determinant) or a lattice basis in periodic arrangements, assuming that it does not contain chaotic perturbations.
In our problem, various definitions may define ${\cal N}_{{\rm pc}}$ relative to the eigenstates $|n\rangle$,
\begin{equation}
{\cal N}^{\{n\}}_{\rm pc}(\Psi_a)=\left (\sum_n |\langle n|a\rangle |^4 \right )^{-1},
\label{eq:35}
\end{equation}
or relative to the basis of a certain observable,
\begin{equation}
{\cal N}^{\{a\}}_{\rm pc}(\Psi_n)=\left (\sum_a |\langle a|n\rangle |^4 \right )^{-1}.
\label{eq:36}
\end{equation}
Both expressions are time independent.
${\cal N}_{{\rm pc}}$ of a state with a certain symmetry can be grossly exaggerated in a basis constructed without regard to this symmetry (for example, in the case of a simple Slater determinant in the nuclear shell model without correct
angular momentum coupling).
The time-reversal symmetry is another relevant example.
The GOE discussed here assumes time-reversal symmetry which at a given time allows us to select basis
states so that all matrix elements are real.  The orthogonal invariance implies that, for any state $\Psi_a$, the matrix elements
$a_n=\langle n|a\rangle$ are real normally distributed random numbers with a variance given by the inverse dimensionality  $1/{\cal N}.$ Then, with either of the above definitions, ${\cal N}_{{\rm pc}}$ is universally
${\cal N}/{3}.$ However, if the basis is chosen as
complex, which is still perfectly good for problems with time-reversal invariance, the amplitudes $a_n$ are complex numbers where both
real and imaginary parts have normal distribution. This leads to the universal value for ${\cal N}_{{\rm pc}}$ being ${\cal N}/2.$
Aiming at the time evolution, we discuss ${\cal N}_{{\rm pc}}$ relative to basis $|a\rangle$  for an evolving state as a function of time,
\begin{equation}
{\cal N}_{\rm pc}(t)=\left (\sum_a |\langle a|\Psi_{a_0}(t)\rangle|^4 \right )^{-1}.
\label{eq:37}
\end{equation}
With this definition, ${\cal N}_{\rm pc}(0)=1$  as the state at the start has only one component $a_0$. As time 
proceeds, the state spreads out
in Hilbert space, and the corresponding complex amplitudes are
$f_a(t).$ For short times, $\exp(-i\hat{H}t)\approx 1-i\hat{H}t$, and, as in nonstationary perturbation theory, the off-diagonal amplitudes are imaginary, $f_a(t)\propto i H_{aa_0}.$ Then $f_a=f^{(r)}_a+i f^{(i)}_a$ are complex quantities whose real and imaginary parts are subject to the normal distributions with different widths.
In fig. \ref{goe_evo_gab_w} we show the time dependence of the three terms comprising overall normalization
\begin{equation}
|f(t)|^2+\sum_{a} {f^{(r)}_a}^2+\sum_{a} {f^{(i)}_a}^2=1.
\label{eq:38}
\end{equation}
The first term shown by the curve (a) is a single diagonal term that represents the {\sl survival probability}
of the initial state. The numerically obtained curve agrees with the analytical form of eq. \eqref{eq:33}. The second 
curve (b) represents the sum of squared real parts of amplitudes $f_a$ related to the width of the distribution of real parts  
${\cal N} \overline {{f^{(r)2}_a}}.$ The curve (c) shows the sum of imaginary parts squared. At any time all three quantities add up to unity. The actual distribution of real and imaginary parts of $f_a$ is shown for $\tau=1$ in inset. This time is labeled on the main plot with a vertical grid line.
Analytical evaluation of distribution widths for real and imaginary parts gives (here $f(t)$ is real due to the symmetry of GOE spectrum)
\begin{equation}
\overline {{f^{(r)2}_a}}=\frac{1}{\cal N} \left (\frac{1}{2} + \frac{f(2t)}{2} - f(t)^2 \right),
\label{eq:39}
\end{equation}
\begin{equation}
\overline {{f^{(i)2}_a}}=\frac{1}{\cal N} \left (\frac{1}{2} - \frac{f(2t)}{2} \right).
\label{eq:40}
\end{equation}
These expressions reproduce the behavior shown in fig. \ref{goe_evo_gab_w}.
The time-dependent number of principal components is
\begin{equation}
{\cal N}_{\rm pc}(t)=\left [|f(t)|^4+{\cal N}\,\, \overline {\left (f^{(r)2}_a+{f^{(i)2}_a}\right )^2}  \right ]^{-1},
\label{eq:41}
\end{equation}
or, using the variances in eqs. \eqref{eq:39} and \eqref{eq:40},
\begin{align}
{\cal N}_{\rm pc}(t)&=\Big [|f(t)|^4+\frac{1}{\cal N} (2+3f^4(t)+  \\
& 
f^2(2t)-4f^2(t)-2 f^2(t) f(2t))  \Big ]^{-1}.
\label{eq:42}
\end{align}
This analytical expression reproduces the numerical results, and the accuracy improves for larger  ${\cal N}.$
In this ``ergodic" limit a single realization
becomes a reliable representation of the ensemble average.
Figures  \ref{goe_evo_NPC_M} and  \ref{goe_evo_NPC} show the
time dependence of ${\cal N}_{\rm pc}$  for three cases, ${\cal N}=10^3,\,\, 5\times 10^3,$ and $10^6.$
In the first plot (the macroscopic limit), using a linear scale, we
show the ratio ${\cal N}_{\rm pc}/{\cal N}$ as a function of time. In asymptotics of large time, all curves approach
${\cal N}_{\rm pc}/{\cal N}=0.5$ that, as discussed above, differs from the GOE limit of $1/3.$ However,
this limit is smoothly reached for relatively small systems. In the macroscopic limit, as clear from the
curve corresponding to ${\cal N}=10^6$, for the most part ${\cal N}_{\rm pc}/{\cal N}\rightarrow 0$ except for a set of peaks.
The reason for this behavior is a macroscopically large survival probability $|f(t)|^2$ that in the limit of an infinite system
remains macroscopic even at very large times.  Only at specific points when $f(t)=0$ (the roots of the Bessel
function, $\tau=1.91585,\,\,3.50779,\,\,5.08673,\dots$), ${\cal N}_{\rm pc}/{\cal N}=1/2.$ These discrete points have zero
measure and for that reason ${\cal N}_{\rm pc}$  cannot be considered an extensive quantity. It is possible that this result is 
connected with dimensional loss discussed for nonequilibrium quantum systems in Ref. \cite{kusnezov:1995}.
The following approximate form for eq. \eqref{eq:42} describes well the trend of NPC and shows interplay between the
${\cal N}\rightarrow \infty$ and ${\tau}\rightarrow \infty$ limits
\begin{equation}
{\cal N}_{\rm pc}(t)=\left [|f(t)|^4+\frac{2}{\cal N}  \right ]^{-1}.
\label{eq:43}
\end{equation}
As long as $f(t)\ne0$
\begin{equation}
\lim_{{\cal N}\rightarrow \infty}{\cal N}_{\rm pc}(t)=|f(t)|^{-4}.
\label{eq:44}
\end{equation}
Considering the asymptotic behavior of the Bessel function for $\tau\gg 1$ at a series of the minima
points of ${\cal N}_{\rm pc}$, we find
\begin{equation}
|f(\tau)|^2 \simeq 1/(\pi \tau^3).
\label{eq:45}
\end{equation}
Thus, for large systems
and for times $\tau^6\ll {\cal N}$ we observe a power law ${\cal N}_{\rm pc}(\tau)\simeq \pi^2 \tau^6$ that
reflects the power-law decay of the survival probability in eq. \eqref{eq:45}.  For extremely long times, those
that on the energy scale are commensurate  with the level spacing $t_{\rm s}\sim N/\lambda$ the ${\cal N}/2$ result is recovered. This is an analog of the known Weisskopf time of the wave packet feeling the quantized grid of discrete levels
\cite{FW}.
The nearly universal behavior for short times is illustrated by Figure \ref{goe_evo_NPC}, where the ${\cal N}_{\rm pc}$ is shown with the use of a log scale. In this limit the survival probability is large, and  the ${\cal N}_{\rm pc}$ is again given by $|f(t)|^{-4}.$
This universal behavior is valid for times $\tau$ less than about 1.5 and while ${\cal N}_{\rm pc}$ is less than about 1000 or ${\cal N}/2$ for smaller dimensions. Although the exact analytic form is known, a simple approximate expression here is  ${\cal N}_{\rm pc}(\tau)\approx \exp(2\tau^2).$
To summarize,  this study of a fully chaotic limit shows that in a large system the
relaxation behavior and the spread of the wave function in Hilbert space are fully determined by the survival probability $|f(t)|^2$
which is the only macroscopic quantity unambiguously defined. Similarly to the radioactive decay of an unstable state, the survival probability has the power-law behavior
at large times. At short times, the most interesting for the processes of the time evolution of observables, the time-dependent behavior, being determined by the survival probability depends on the density of states at the most remote ends of the spectrum.
\section{Towards realistic systems}
\subsection{Random two-body interactions}
\begin{figure}[h]
\begin{center}
\includegraphics[width=0.99\linewidth]{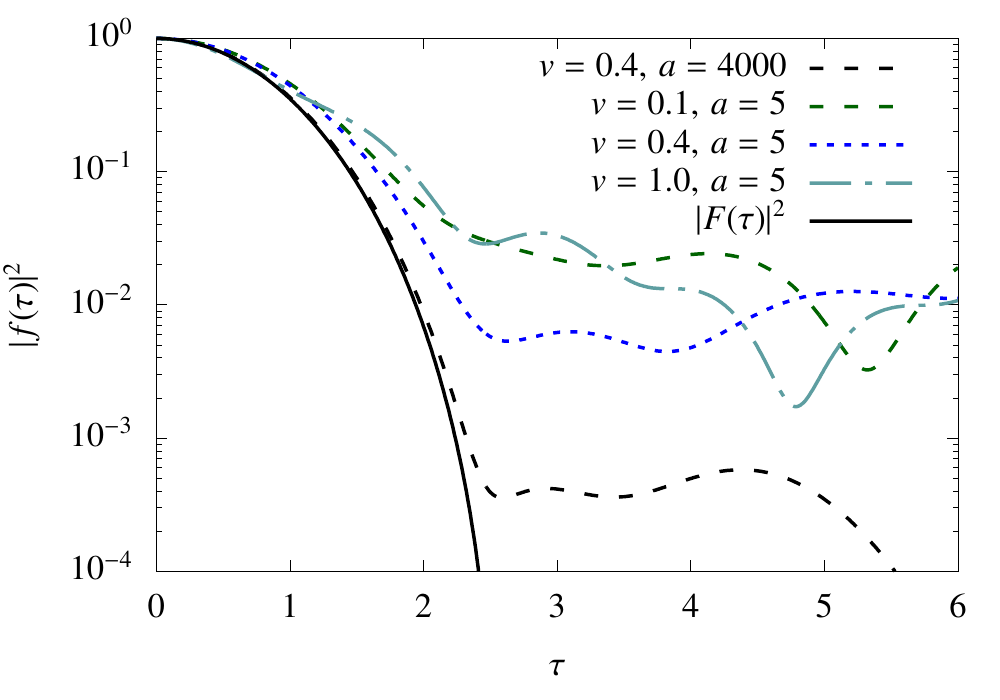}
\end{center}
\caption{\label{bos_ff} The dependence of the survival probability for several states in the interacting boson model compared
to the averaged function $|F(\tau)|^2.$}
\end{figure}

\begin{figure}[h]
\begin{center}
\includegraphics[width=0.99\linewidth]{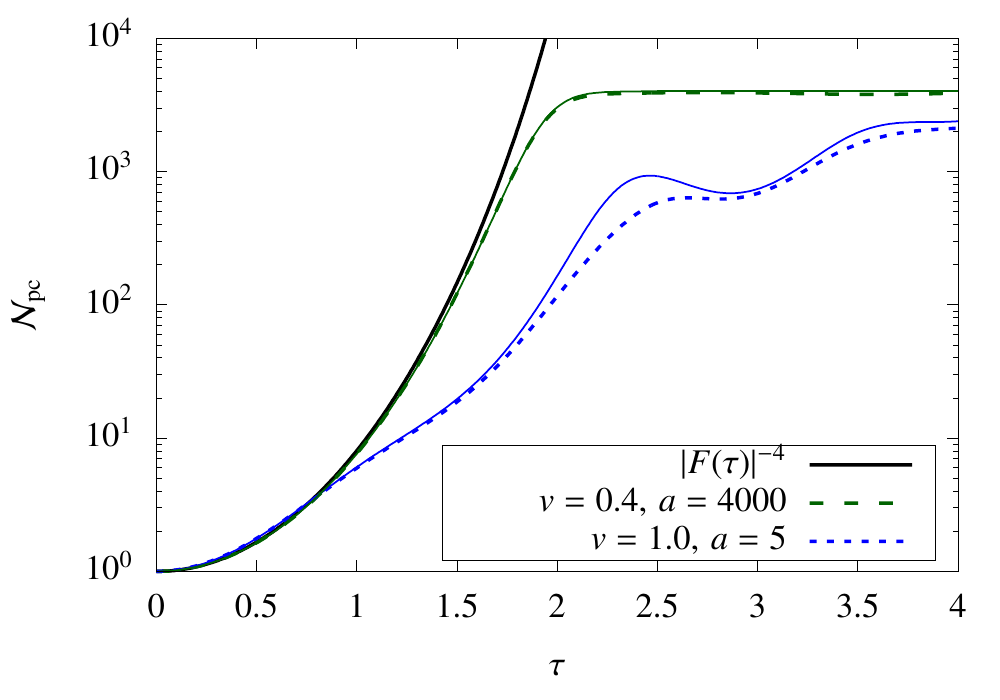}
\end{center}
\caption{\label{bos_NPC} The ${\cal N}_{\rm pc}$ as a function of time for two representative states in the interacting boson model,   compared to $|F(\tau)|^{-2}.$  The thin solid lines show approximations given by eq. \eqref{eq:43}. }
\end{figure}

The
discussed above
chaotic limit implies an extreme assumption that the matrix elements of the interaction between any two
many-body states
are statistically identical; the system is invariant under orthogonal
basis transformations.
The realistic many-body systems usually have the mean-field basis that is special and thus violates
orthogonal invariance.
There is commonly a hierarchy of dynamics: mean-field, two-body interactions,
and interactions of higher order which are often ignored or partly included in average into lower order terms.
The practical Hamiltonian matrices are very sparse. With the standard two-body interaction, the same elements
are repeated in the matrix being identical on different backgrounds of other spectator particles.
The two-body random ensembles have been extensively studied in the literature.
For sufficiently strong interactions, and/or higher in the excitation spectrum, the components of wave functions in the
original basis
approach the normal distribution. This leads to the Porter-Thomas distribution of reduced widths which is a
robust result reflecting complexity of many-body states. Therefore, many of our results discussed above remain
valid (with some limitations).
The survival probability for each individual state can be different,
especially for the states at the edges of the spectrum or in the case of
weak interaction. We can no longer assume $f(t)=F(t).$ The survival probability is a Fourier transform of the strength function widely discussed in the literature \cite{BISZ}.
In Fig. \ref{bos_ff} we show $|F(\tau)|^2$ and several representative examples of the survival probability $|f(\tau)|^2$ as
a function of the scaled time $\tau$ for a model of six spinless bosons that can occupy eleven single-particle levels
with equidistant level spacings taken as a unit of energy. The two-body interactions are selected at random
in a GOE-like way with the strength parameter $v.$ Results for a single realization are shown
with no averaging over an ensemble of such systems. The underlying mechanism is universal resulting from the
effectively strong mixing of simple states at their high density emerging by combinatorial reasons.
For short times $\tau<1$ (the most physically relevant region for decay $|F(\tau)|^2$) the result is universal,
$|F(\tau)|^2=\exp(-\tau^2)$, reflecting Gaussian tails in the energy spectrum.
For a sufficiently strong two-body interaction,
$|F(\tau)|^2 $, being expressed in scaled time $\tau=\lambda t$, is independent of $v$ and falls off faster at large times.
This curve is shown in  fig. \ref{bos_ff}  with the solid line.
The remaining curves show the survival probability for different initial states $a$ that are eigenstates of the non-interacting system. There are total of
${\cal N}=8008$ many-body states
numbered in such a way that configurations with lower $a$ have more particles in lower orbits,  While survival probabilities look differently on the log scale, the differences only emerge
at remote time when probabilities themselves are small, while at short times $f(\tau)\approx F(\tau).$  This agreement extends further
for complex states in the middle of the spectrum, such as the $a=4000$ state shown with a black dashed line. 
States at the edges
of the spectrum are strongly influenced by the energy cutoff  and, as in realistic decay problems, their evolution slows down at times of the order of the inverse  distance to thresholds in the energy domain.
Figure \ref{bos_NPC} shows ${\cal N}_{\rm pc}$ for the two most distinct states. The results with high accuracy are explained by the discussion in the previous section. The results from the simplified
equation \eqref{eq:43} are already in a good agreement with numerical data. The reason behind this agreement is the normal distribution of components in complex wave functions which is true for states regardless of their position in the spectrum (essentially the central limit theorem). This makes the survival probability nearly an exclusive characteristic of the time evolution.
\subsection{Shell model dynamics}
\begin{figure}[h]
\begin{center}
\includegraphics[width=0.99\linewidth]{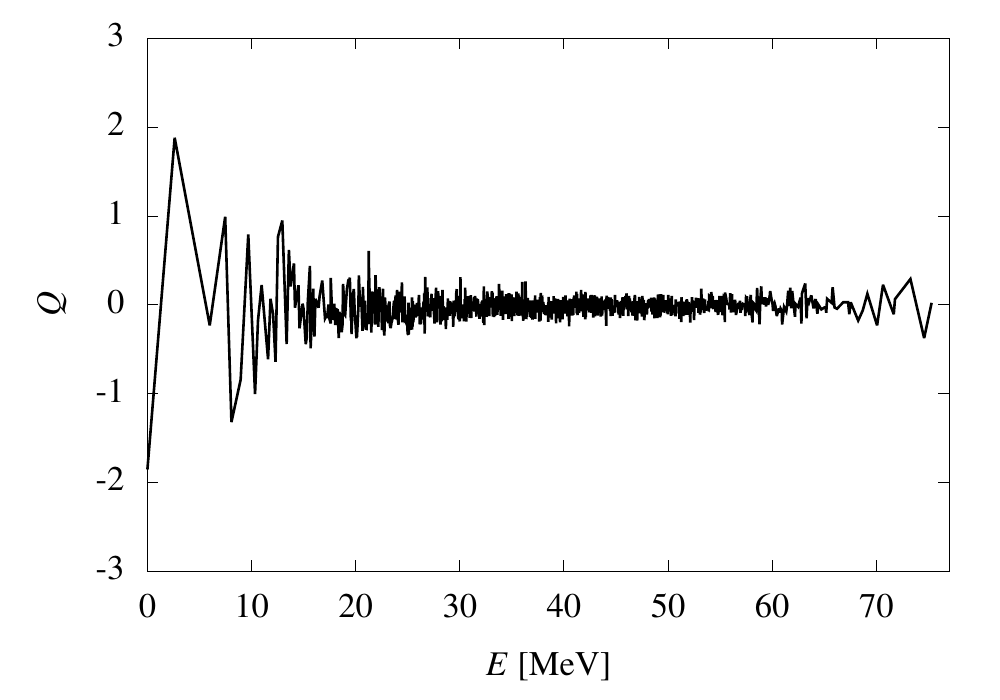}
\end{center}
\caption{\label{fig:mg_e} Quadrupole moment of all states $2^+$ in the $sd$-model
of  $^{24}$Mg as a function of excitation energy. }
\end{figure}

\begin{figure}[h]
\begin{center}
\includegraphics[width=0.99\linewidth]{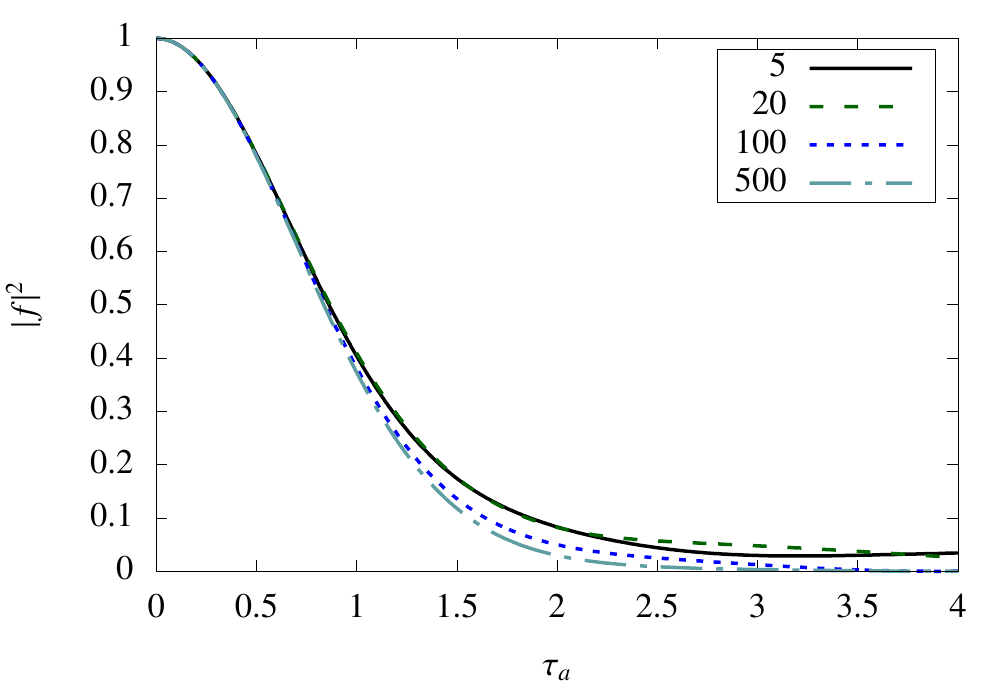}
\end{center}
\caption{\label{fig:mg_evo_f} Survival probability for several states $a.$
}
\end{figure}

\begin{figure}[h]
\begin{center}
\includegraphics[width=0.99\linewidth]{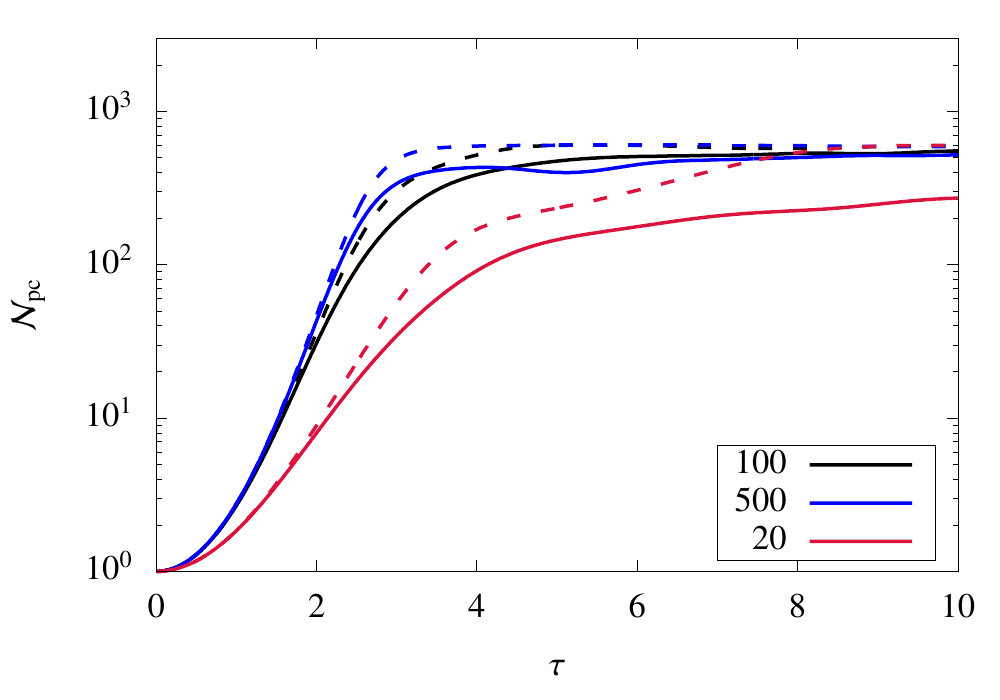}
\end{center}
\caption{\label{fig:mg_evo_NPC} ${\cal N}_{\rm pc}$ as a function of time for several states. Dashed lines show the approximate result obtained using the survival probability, eq. \eqref{eq:43}.
}
\end{figure}

\begin{figure}[h]
\begin{center}
\includegraphics[width=0.99\linewidth]{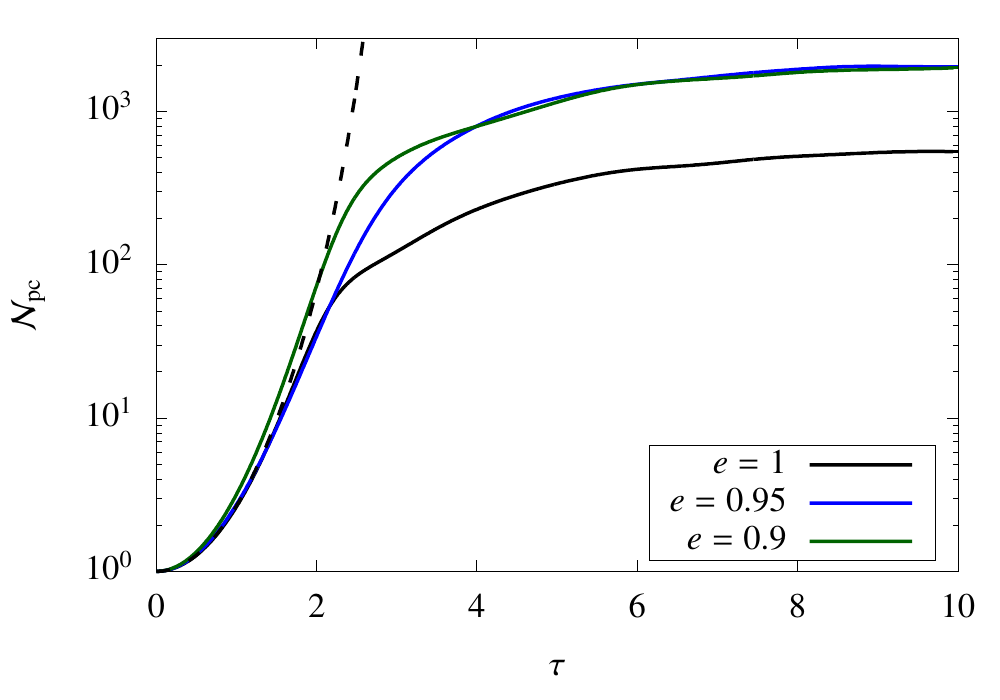}
\end{center}
\caption{\label{fig:mg_evo_NPCT} ${\cal N}_{\rm pc}$ as a function of time for a state $a=100$. Lines correspond to a different quadrupole operator used with a slight variation of the neutron effective charge. Dashed line shows $|f(\tau)|^{-4}.$
}
\end{figure}

\begin{figure}[h]
\begin{center}
\includegraphics[width=0.99\linewidth]{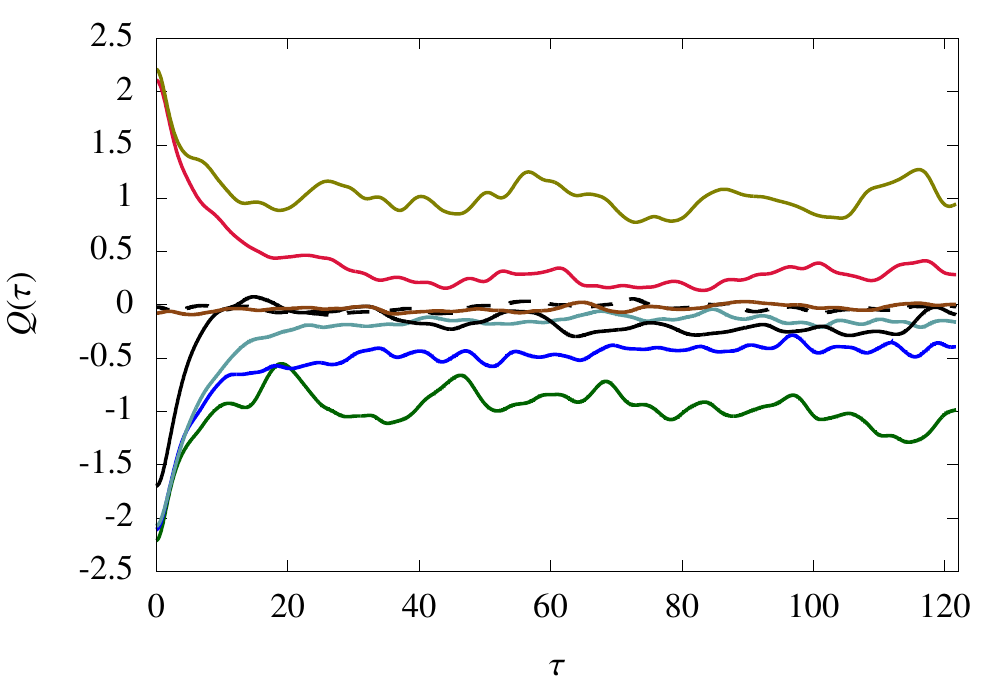}
\end{center}
\caption{\label{fig:mg_evo} Time evolution of ${Q_a}(t)$ for states $a=0,1,2, 20, 600, a_{\rm max}-1, a_{\rm max}.$  
The case $a=0$ represents the most negative value of ${\cal Q}$ that can be measured in this system and $a_{\rm max}$ corresponds to the biggest $Q.$ Dashed line on the plot corresponds to 
a time-evolved randomly picked initial vector where all components have Gaussian distribution with the same mean.
}
\end{figure}

\begin{figure}[h]
\begin{center}
\includegraphics[width=0.99\linewidth]{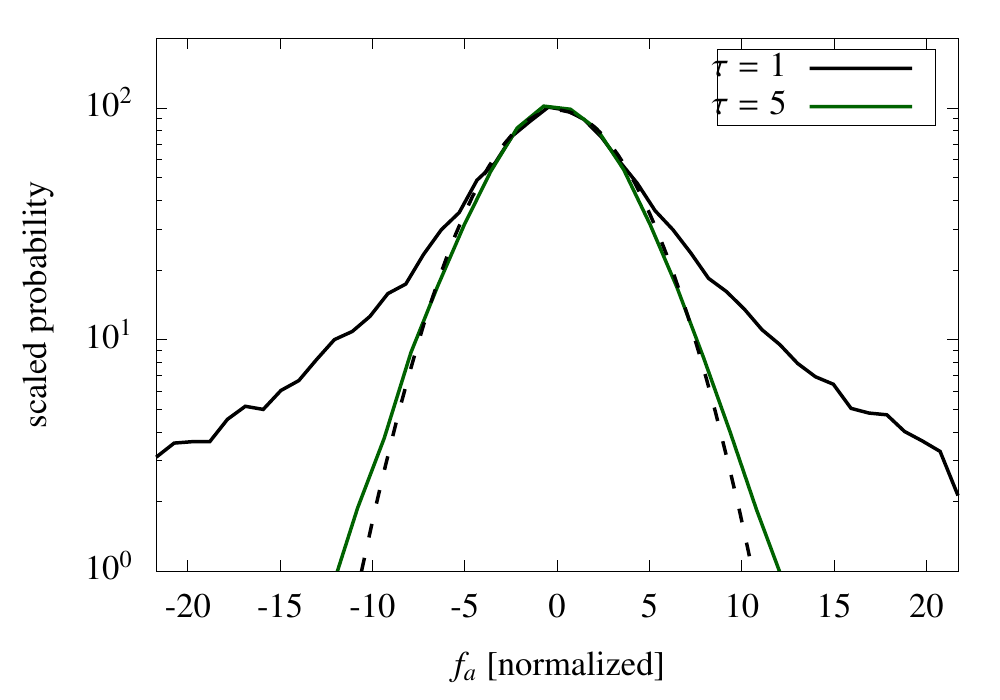}
\end{center}
\caption{\label{fig:mg_q_hist} Distribution of $f_a^{(i)}$ which are the imaginary parts of off-diagonal matrix elements of the propagator
for two times $\tau=1$ and $\tau=5.$ Axis are normalized so that all curves have the same position of the peak and the same curvature at the peak. The dashed line shows a normal distribution.
}
\end{figure}

\begin{figure}[h]
\begin{center}
\includegraphics[width=0.95\linewidth]{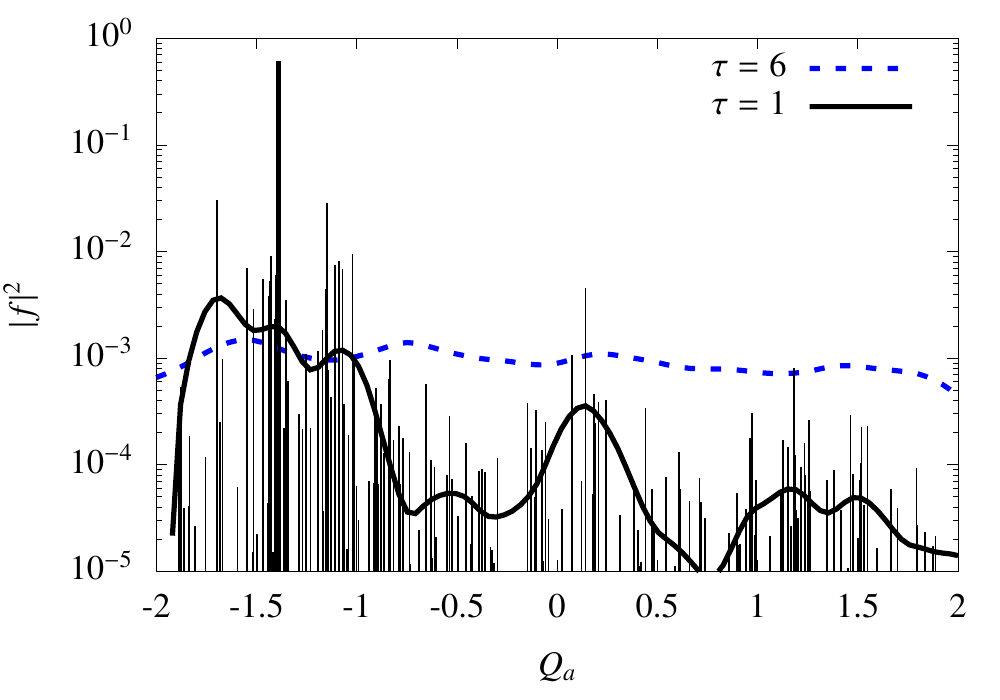}
\end{center}
\caption{\label{fig:mg_q_sin} Distribution of amplitudes squared  $|f_{aa_0}|^2$ for initial state $a_0=100$
as a function of $Q_a$ for two times $\tau=1$ and $\tau=6.$
}
\end{figure}

Here we look at the time evolution described by a full quantum machinery of the nuclear shell model ({\sl configuration interaction}). Various shell-model versions are widely used
by nuclear practitioners as the most reliable detailed framework for description of low-lying nuclear states
and reactions.
For our purpose it is not important if the model contains a fixed
inert core of occupied orbitals that is not destroyed by low-lying dynamics, or starts ``{\sl ab initio}" from
the vacuum state.
The effective Hamiltonian is usually found in
a combination of bare nucleon-nucleon interaction and phenomenological corrections reflecting the renormalization
of this interaction by the medium effects, as well as by the fact of the space truncation. Successful Hamiltonians
describe quite reliably (below excitation energy limited by space truncation) the properties of the excited
nuclear states, electromagnetic and weak transitions, level density, and reaction probabilities.
The nuclear stationary states in the vicinity of the ground state typically can be understood in terms
of occupancies of low-lying single-particle orbits according to the Fermi statistics of nucleons. The most
coherent parts of the interaction create effects of superfluid pairing and collective branches of the excitation
spectrum. With increase of excitation energy and combinatorial level density, the wave functions
of stationary states become more and more complicated combinations of simple fermionic configurations. This
leads to the fast growth of the number of principal components
in the expansion of a stationary wave function in terms of basis partitions. Both the energy spectrum
in a class of states with given values of exact constant of motion (total nuclear spin and, in a good
approximation, parity and isospin) and properties of the stationary states, including the level spacing statistics
and the so-called $\Delta_{3}$ statistics of level sequences, approach the main GOE predictions
\cite{porter,brody,annu}. Such properties are smoothly changing along the energy spectrum revealing thermodynamic
behavior \cite{big}. Here one can expect a characteristic time dynamics of non-stationary states initially
selected as a result of various ``prehistoric" processes.
The stage of nuclear dynamics we try to describe is essentially thermalization with no external heat bath.
This is similar to what is historically called the compound-nucleus formation \cite{FW}. The realistic process starts with
a simple excitation, usually of only few degrees of freedom. For a complex nucleus it can be absorption
of a neutron in an astronomic process, preceding nuclear reaction, or electromagnetic excitation. This non-stationary 
{\sl doorway state} \cite{AZ07} located in the region of sufficiently high level density is the wave packet of many stationary functions with certain initial phases corresponding to a simple mode of excitation. In this sense it reminds an initial state of a classically chaotic system
with many degrees of freedom. Then a natural question is the evolution of such initial excitations, the prototype
of which in classical chaotic dynamics typically reveals exponential divergence in the process of evolution.
Here we are looking at time intervals of statistical equilibration which are still smaller than the life time
with respect to the irreversible decay into continuum.
Attempts to theoretically describe the time evolution of a quantum many-body system
are frequently based on the sequence of processes
involving transitions between different classes of excited states characterized by a number of particle-hole
excitations \cite{agassi} counted from the non-interacting ground state, see also \cite{BISZ} and references therein. 
Such processes traditionally can be described by a tree-branching population dynamics \cite{BIS19}
as in a real chain of radioactive transformations.
Below we give an example of the exact solution of time-dependent quantum dynamics avoiding assumptions
of the tree dynamics where the motion is essentially assumed to proceed only in the forward direction
due to the higher dimension of every next class of states. We will see that, in a realistic chaotic many-body
dynamics already started from the non-stationary state with all interactions present, this assumption does not work 
(except maybe for special models like spin systems with interaction
of close neighbors). After an initial stage, the system enters the region where the states have a similar
population and the thermalized motion does not have a certain direction.
Our example will be the shell-model description of the nucleus $^{24}_{12}$Mg$_{12}$ well studied
experimentally and theoretically. In the simplest $sd$ shell-model we have four neutrons
and four protons occupying $1d_{5/2}, 2s_{1/2}$ and $1d_{3/2}$ single-particle orbitals on top of
the inert core of $^{16}$O. This system is sensitive to the quadrupole deformation, and our variable
$\hat{Q}$ will be the nuclear quadrupole moment $\hat{Q}_{20}$, the tensor component with angular momentum
$L=2$ and its projection $M=0$.  Under the assumption of equal effective
charges for neutrons and protons which would make the operator $\hat{Q}$ represent the nuclear density, the Hamiltonian dynamics is
confined to the subspace of spin $J=2$
and zero isospin, conserved by the Hamiltonian. The model space contains ${\cal N}=1206$ such states;
if isospin is allowed to be broken with angular momentum being strictly conserved, the number of states involved grows to ${\cal N}=4514,$ which are all states with spin-parity
quantum numbers $2^{+}.$
The harmonic oscillator single-particle wave functions and the USDB effective interaction \cite{brown:2006} are used to evaluate the quadrupole moment expressed in the units of the oscillator length squared.  The operator $\hat{Q}$  is defined in spherical basis so it is different
from the Cartesian operator by a factor of $\sqrt{16\pi/5}$. The lowest stationary states in $^{24}$Mg are known
to have a pronounced trend to deformation and the corresponding rotational band. The quadrupole moment of
the actual first $2^+$ stationary state is $Q_0=-1.853$ indicating significant collectivity. In Figure \ref{fig:mg_e},
the quadrupole moments of all  stationary $2^+$ states in $^{24}$Mg are shown. The average and the variance for the quadrupole moment operator are
\begin{equation}
\overline{Q}=\frac{1}{\cal N} {\rm Tr} \hat Q= -0.021, \quad \frac{1}{\cal N} {\rm Tr} (\hat Q-\overline{Q})^2= 1.08^2
\label{eq:46}
\end{equation}
The developing state is initialized as an eigenstate $|\Psi_{a_0}(0)\rangle = |a_0\rangle$ of the quadrupole operator,
\begin{equation}
\hat{Q} |a_0\rangle =Q_{a_0} |a_0\rangle.                                                        \label{eq:47}
\end{equation}
In this model space the eigenvalues $Q_a$ of the operator $\hat{Q}$ vary between the lowest $Q_{0}=-2.209$ and
the highest one $Q_{{\rm max}}=+2.209$ which we order with the increasing subscript assuming $Q_{a}<Q_{a'}$
if $a<a'$.
Starting with this initial state at $t=0$, we look at its dynamics for $t>0,$
\begin{equation}
|\Psi_{a_0}(t)\rangle=e^{-i\hat{H}t}|a_0\rangle,                   \label{eq:48}
\end{equation}
where the expectation value changes as in Eq. \eqref{eq:4}.
The state $|\Psi_{a_0}(t>0)\rangle$ is not an eigenstate of the quadrupole operator anymore.
The survival probability for different states $|a\rangle$ is shown in Fig. \ref{fig:mg_evo_f}.  Apart from
long-time tails, the curves show remarkable identity reflecting similarity in the strength functions
of the starting states.
The time scale can be examined from the average spectral width,
\begin{equation}
\lambda^2=\frac{1}{{\cal N}} {\rm Tr} (\hat{H}-\overline{E})^2 ,\quad {\rm where} \quad \overline{E}=
\frac{1}{{\cal N}} {\rm Tr} \hat{H}.                                                                    \label{eq:49}
\end{equation}
For the $2^{+}$ states generated by the USDB Hamiltonian, this second moment is $\lambda=12.17$ MeV.
The energy spread of different initial states $|a\rangle$,
\begin{equation}
\lambda_a^{2}=\langle a|\hat{H}^2|a\rangle- \langle a|\hat{H}|a\rangle^2,                      \label{eq:50}
\end{equation}
is generally smaller but of the same order, $\lambda_a=9.0$, 11.3, 6.9, 8.7, 9.3, and 8.7  MeV for $a=0,$ 1, 20, 100,400, and
500,  respectively.
Here the mean field plays an important role lowering the average value of
$\lambda_a$ compared to $\lambda$,
\begin{equation}
\overline{\lambda^2_a}=\lambda^2-{\rm Var}[H_{aa}].
\label{eq:51}
\end{equation}
where
\begin{equation}
{\rm Var}[H_{aa}]=\frac{1}{\cal N}\sum_a H^2_{aa} - \left (\frac{1}{\cal N}\sum_a H_{aa} \right )^2.
\label{eq:52}
\end{equation}
For the high-dimensional GOE, the diagonal matrix elements have variance $2\lambda/{\cal N}$, and
this difference can be ignored. Correspondingly, the diagonal matrix elements are negligible in the GOE case.
For shell model matrices and few-body operators, the diagonal matrix element squared can be comparable to the sum of all off-diagonal matrix elements squared, and thus the difference is significant.
For the quadrupole operator in $^{24}$Mg, $\sqrt{\overline{\lambda^2_a}}=9.83$ MeV.
The fluctuations around average are small and lead to small variations in the width of the survival probability curve. These variations completely disappear if the adjusted timescale $\tau_a=\lambda_a t$ is used for each of the curves as shown in Fig.
\ref{fig:mg_evo_f}.
The ${\cal N}_{\rm pc}$ is addressed in Figure \ref{fig:mg_evo_NPC}. For most of the states, this characteristic is universal and agrees well with the approximation of eq. \eqref{eq:43}. We also show state $a=20$ to
 illustrate the extreme deviation from typical. Despite this deviation, it is still well described by eq. \eqref{eq:43}.
Overall, the survival probability and ${\cal N}_{\rm pc}$ demonstrate chaotic properties consistent
with the previously discussed chaotic models.
The lower levels may exhibit fluctuations and deviations in dynamics at remote times but some of those aspects may be of no particular physical significance.
For example, ${\cal N}_{\rm pc}$, in addition to previously discussed issues, is also sensitive
to symmetry conservation and its weak violation. In fig. \ref{fig:mg_evo_NPCT} for the state $a=100$,
${\cal N}_{\rm pc}$ for the original isospin conserving quadrupole operator with effective charges $e_n=e_p=1$ is compared with the ones where $e_n$ is slightly lowered.
This causes sharp discontinuity at around $\tau=2.5$ as the number of available states grows from 1206  to 4514. Thus similar to our examination of the GOE limit, the ${\cal N}_{\rm pc}$ result is universal and well defined only in the limit where it is fully determined by the survival probability \eqref{eq:44}
shown with the dashed line.
Figure \ref{fig:mg_evo}
shows the time evolution of the quadrupole moment for several initial states $|a\rangle$. With increasing time, the expectation values $Q(t)$
quickly shrink towards the $Q=\overline{Q}\approx 0$ value, with the exception of the most extreme cases at the edges, $a=0,1,2$, and at the opposite end around $a=a_{\rm max}$. For the majority of states, the mean values $Q(\tau)$ for $\tau>10$ fluctuate around zero. The survival of non-zero mean values for the lowest and
highest $\Psi_a(t)$ is related to quadrupole collectivity in the USDB Hamiltonian that allows this coherence to survive
in the evolution. The remaining states for $\tau>10$ are statistically indistinguishable from the random state shown
with the dashed line.
The few lowest states, especially $a=0$ and $a=1$, retain a noticeable mean quadrupole moment
even after a long evolution. This is a typical behavior of collective modes,
like {\sl giant resonances} in nuclei, which are not stationary states but doorways to mixing (damping) with
numerous compound states \cite{AZ07,ZS}. Nevertheless, the collective strength being spread over many underlying
stationary states still is observable being concentrated in a certain spectral region. This is an analog of
{\sl scars} known in classical chaotic systems \cite{KH}.
\begin{figure}[h]
\begin{center}
\includegraphics[width=0.99\linewidth]{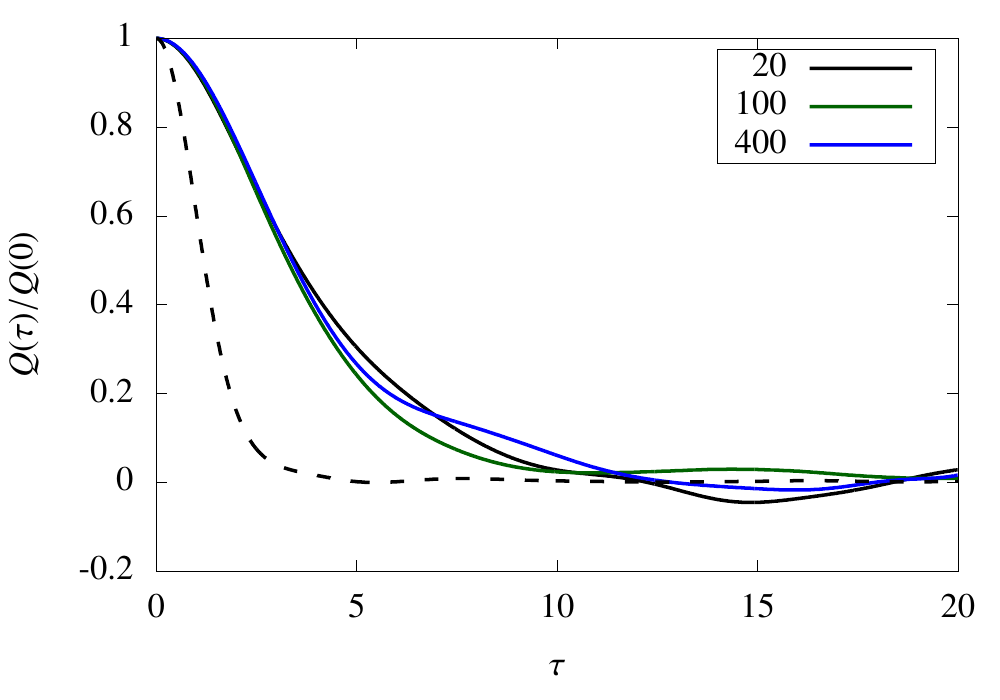}
\end{center}
\caption{\label{fig:mg_evo_q}
Time evolution of $Q(t)/Q(0)$ for several initial states, $a_0=20, 100,$ and 400. The typical form of the survival
probability $|f(\tau)|^2$ is shown by the dashed line. �
}
\end{figure}
It is remarkable that even the higher $a$-states reach statistical
values on a significantly longer time scale than for the survival probability.
In Fig. \ref{fig:mg_evo_q} we study the short-time evolution of the variable $Q$. With the exception of
low-lying collective states, the time evolution of the quadrupole moment expressed relative to its value at $t=0$ is universal. However, while its behavior is similar to that of the survival probability, the life time is significantly (about a factor of three) longer and eq. \eqref{eq:26} is not valid. The dashed line shows representative $|f(\tau)|^2$, that,
as concluded from Fig. \ref{fig:mg_evo_f}, is nearly the same for all these states. An extended lifetime of  perturbations
appears to be typical for many-body systems where single-particle degrees of freedom and mean field lead to correlations between the Hamiltonian and operator $\hat{Q}$ dominated by diagonal matrix elements. These correlations
exhibit themselves in two aspects where the discussion that led to equation  \eqref{eq:26} needs to be modified.
It is appropriate to draw attention here to the non-chaotic limit of the squeezed state of the harmonic oscillator
mentioned in Section 2.2. In that case the
extremely fast decay of the survival probability is also decoupled from the time evolution of the coordinate.
First, the distribution of the off-diagonal matrix elements $f_{aa'}$ depends on the Hamiltonian in the basis of operator $\hat{Q}.$ Consider for example the short-time limit for $a\ne a'$:
\begin{equation}
|f_{a a'}(t)|^2 =t^2 \left ( \langle a|\hat{H}|a'\rangle^2-\langle a|\hat{H}^2|a'\rangle\right )
\label{eq:53}
\end{equation}
In the limit of ensemble average, the second term does not contribute and we are left with the square of the
off-diagonal matrix element   $H_{aa'}=\langle a|\hat{H}|a'\rangle.$ In contrast to the uncorrelated limit,
the matrix $H_{aa'}$
is a diagonal-dominated banded matrix.
In fig. \ref{fig:mg_q_hist}  we show the distribution of imaginary parts 
$f_a^{(i)}$ for two times, $\tau=1$ and $\tau=5$, and compare them with the Gaussian shown with a dashed line.
Given our interest in the form of the distribution, we scale the curves making their peaks coincide and
normalize $f_a^{(i)}$
so that the curvatures at the peak coincide. The probabilities in a logarithmic scale show that for short times the distribution is more of the Lorentzian-type with exponentially long tails.
The deviations from the normal distribution appear at short-time scales
within about the same time interval where the survival probability is large.
The effect of long tails on ${\cal N}_{\rm pc}$ is minimal although, with more in-depth examination of fig. \ref{fig:mg_evo_NPC}, one could notice that for times $\tau$ between about 1 and 6 the predicted ${\cal N}_{\rm pc}$
is systematically higher.
The second circumstance needed to explain the extended survival of quadrupole deformation seen in fig. \ref{fig:mg_evo_q} is the correlation between large off-diagonal amplitudes $f_{aa'}$ and similar values of the quadrupole moments
$Q_a$ and $Q_{a'}.$
At short times, we find  from eq. \eqref{eq:5}
\begin{equation}
Q_a(t)=Q_a(0)-\frac{t^2}{2} \langle a| [\hat{H}, [\hat{H},\hat{Q}]]|a\rangle.
\label{eq:54}
\end{equation}
For a more general picture we can examine the sum
\begin{equation}
Q_a(t)=Q_a |f_{aa}(t)|^2 + \sum_{a'\ne a} Q_{a'} |f_{a'a}(t)|^2.
\label{eq:55}
\end{equation}
Without any particular correlation, as in the GOE limit, the summation over all off-diagonal terms weighted by
random amplitudes averages out to the time-independent value $\overline{Q}$ which is nearly zero in this model.
The situation in $^{24}$Mg at short times is different as shown in Fig. \ref{fig:mg_q_sin} where the distribution of amplitudes
$|f_{aa_0}|^2$ with fixed $a_0=100$ is shown as a function of $Q_a$ at two different times. The plot represents
the time-dependent strength function
\begin{equation}
G_{a_0}(t,Q)=\langle a_0|e^{i\hat{H}t} \delta (Q-\hat{Q}) e^{-i\hat{H}t}|a_0\rangle
\label{eq:56}
\end{equation}
which is shown at two instances, $\tau=1$ and $\tau=6.$
For $\tau=1$, black impulse lines show the actual distribution that includes the diagonal survival probability as the highest
peak nearly two orders of magnitude higher than any off-diagonal amplitude.  The solid continuous line shows the smoothed
average of off-diagonal $|f_{a}|^2$. The non-uniform distribution clearly puts more weight to values of the quadrupole moment
close to  $Q_{a_0}$ which explains why the $Q(t)$ remains significant for much longer time than the survival probability, while 
the distribution becomes nearly uniform as seen at  $\tau=6$ where only the smoothed average is
shown.
\section{Conclusions}
In this work we studied the time evolution of observables and other characteristics of the wave functions,
such as the number of principal components, in closed quantum systems that range from chaotic to regular.
Here one can recall a popular quote from Atkins \cite{atkins}, {\sl We are the children of chaos, and the deep structure
of change is decay.}
The decay of initial perturbations is an important element of the quantum many-body dynamics which is closely tied to questions
of thermalization in mesoscopic systems, connections between classical and quantum complexity as well as to the general
physics of time-dependence and decay.
In the chaotic limit represented by the Gaussian Orthogonal Ensemble we put forward an asymptotically exact 
analytic expressions that describe the universal time evolution of a generic initial state, its spread in the Hilbert 
space and the corresponding number of principal components which can be considered as a quantum counterpart
for the classical propagation in the phase space. We show that, in this special limit of orthogonal invariance, 
the time evolution is defined exclusively by the survival probability of the initial wave function.
The chaotic limit of GOE sets the stage for the analysis of more realistic quantum many-body systems, where it is 
shown that the survival probability remains an important basis-independent generic characteristic of complex wave 
functions. 
As known from multiple studies of quantum-mechanical decay \cite{decay}, such processes reveal certain typical
characteristics 
reflected in the time dependence of the survival probability. The early stage of the evolution, given by the $t^2$ 
dependence is determined by the quantum unitarity. It is generally followed by the exponential decay that
in the long time limit switches to a power-law $\sim 1/t^\beta$ where the exact value of $\beta$ depends on 
the asymptotics of the level density at the edges of the spectrum. Above, see Fig. \ref{fig:mg_evo_f}, one can
roughly identify these limits, but, for the states we considered, the interval of exponential decay is narrow while it is 
typically the main observable stage of radioactive decay. We have found that more realistic systems with the 
skeleton of the regular mean-field show an extended decay of perturbations, much longer than the time scale 
given by the survival probability. We have shown also that the dynamics typically cannot be described by the tree
branching  with several intermediate stages involving every time a new class of states (although such models are 
certainly possible). The end of the characteristic time evolution allows, instead of the irreversible decay to 
continuum, as in the radioactive decay, just thermal-type fluctuations on the level determined by the mean
energy stored in the initial non-equilibrium state.
We did not find any quantum manifestation of the classical Lyapunov exponent.  One can argue that classical 
sensitivity to initial conditions is diminished and smeared by the quantum-mechanical uncertainty. In the language 
of the Feynman path integral formulation, in a realistic many-body system, the slightly modified classical paths that exponentially diverge classically have nearly identical action and become smoothly averaged in quantum mechanics.  
This claim refers to finite systems where, in contrast to their macroscopic limit, the interaction does not immediately
induce phase transformations as in the case of Fermi-gas instability with respect to Cooper pairing. One supporting 
example for such arguments is a class of problems where Heisenberg operators satisfy classical equations of motion. 
The chaotic nature of those equations would not manifest itself for different initial conditions that simply correspond to different wave functions chosen to evaluate expectation values of the Heisenberg operators. Finally, wave function 
components of the isolated quantum systems in the finite Hilbert space  that we considered have classical quasi-periodic dynamics \cite{weinberg:1989,izrailev90}. It was stressed long ago \cite{chirikov92} by Chirikov that 
{\sl The quantum chaos is 
finite-time statistical relaxation in discrete spectrum}. At the same time Wikipedia gives the following definition: 
{\sl Quantum chaos is a branch of physics which studies how chaotic classical dynamical systems can be described in 
terms of quantum theory. The primary question that quantum chaos seeks to answer is: ``What is the relationship between quantum mechanics and classical chaos?"}. This statement is extremely restrictive - currently we are interested in physics of realistic systems, mainly many-body, which do not have a clear classical limit but reveal clear and practically important 
quantum chaotic phenomena, such as thermalization, decoherence, stability and phase transitions,  which still require a lot of work for their understanding.   
\section*{Acknowledgments}
This material is based upon work supported by the U.S. Department of Energy Office of
Science, Office of Nuclear Physics under Award Number DE-SC-0009883. We thank N. Auerbach for
useful discussions at the initial stage of the work. We are also grateful to Pierre Nzabahimana for his
participation in numerical simulations.

\end{document}